\documentclass[runningheads]{svmult}

\usepackage{graphicx}  
\makeindex             

\begin{document}
\title*{X-ray transient galaxies and AGN}
%
%
%
%
%
\author{Dirk Grupe\inst{1}
}
%
%
%
\institute{MPI f\"ur extraterrestrische Physik, PO Box 1312, D-85741 Garching,
Germany
}

\maketitle              

\begin{abstract}
X-ray transience is the most extreme form of variability observed in AGN or
normal in-active galaxies. While factors of 2-3 on timescales of days to years
are quite commen among AGN, X-ray transients appear only once and vanish from
the X-ray sky years later. The ROSAT All-Sky Survey was the tool to discover
these sources. X-ray transience in AGN or galaxies can be caused by dramatic
changes in the accretion rate of the central black hole or by changes of the
properties of the accretion disk.
\end{abstract}

\section{Introduction}
Our sample of soft X-ray AGN contains 113 sources selected from the
ROSAT All-Sky Survey (RASS, \cite{voges1999}) by
the PSPC count rates CR $>$ 0.5 cts
s$^{-1}$ and the hardness ratio HR $<$0.0. Pointed PSPC and HRI observations are
available for 60 and 50 sources, respectively. All in all, for more than 80 
sources at least one pointed observation is available \cite{grupe2001a}.
In this way we have a
tool to search for long-term large amplitude variations. Fig. 1 displays the
RASS vs. pointed observation count rates. HRI count rates have been converted
into PSPC count rates assuming no spectral change between both observations.
The solid line marks no change, the short-dashed line a change by a factor of 10
and the long-dashed line by a factor of 100 between the RASS and the pointed
observation. Four sources turned out to vary by factors of almost 100 or even
more: {\bf WPVS 007, IC 3599, RX J1624.9+7554}, and {\bf RX J2217.9--5941}. The
first three are X-ray transients while  RX J2217.9--5941 is a possible
X-ray transient candidate.

\begin{figure}[b]
\begin{center}
\includegraphics[width=.7\textwidth]{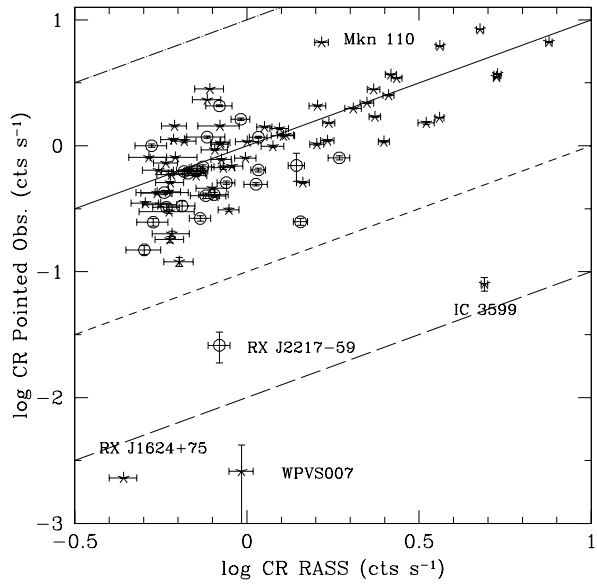}
\end{center}
\caption[]{RASS vs. pointed observation count rates}
\label{rass}
\end{figure}

\section{X-ray transient sources}

\subsection{WPVS 007}

The Narrow-Line Seyfert 1 galaxy (NLS1)
WPVS 007 was {\it the} softest AGN observed during the RASS \cite{grupe1995b} 
and had a mean count rate of about 1 cts s$^{-1}$. In
later pointings the source shows only the flux expected from a normal inactive
galaxy. 
A possible scenario to explain this dramatic turn-off is a temperature
change in the Comptonization layer of the accretion disk that shifts the soft
X-ray photons out of the ROSAT energy window.

\subsection{IC 3599}

The Seyfert 2 galaxy IC 3599 has shown an X-ray outburst during the RASS
followed by a response in its optical emission lines (\cite{grupe1995a}, 
\cite{brandt1995}). A possible explanation of this X-ray outburst is an accretion
event either caused
by an instability of the accretion disk or by a tidal disruption of
a star orbiting around the central black hole.

\subsection{RX J1624.9+7554}

RX J1624.9+7554 has shown a dramatic derease of its X-ray flux by at
least a factor of more than 200 between the RASS and a pointed observation 1.5
years later (Grupe et al. 1999). Optical spectroscopy identified this source as
a normal in-active galaxy. The most plausible explanation for this X-ray event
is the tidal disruption of a star by the central black hole. Other in-active
galaxies in which an X-ray outburst have occured are
NGC 5905, 
RX J1242.6--1119 (\cite{kombade},
\cite{komossa1999}; see also S. Komossa's contribution
in these proceedings), and RX J1420.4+5334 \cite{greiner2000}.

\subsection{RX J2217.9--5941}

The NLS1 
RX J2217.9--5941 is a possible X-ray transient candidate. It is highly variable on
time scales of days as well as years \cite{grupe2001b}. During its two-day
RASS observation the count rate decrease by a factor of 15. Observed several
times in pointed observations by ROSAT and ASCA the source has become fainter
over the years.
It is not clear yet if this source will become a transient. However, due to the
black hole mass of $\approx$ 10$^8$$M_{\odot}$ the timescales are larger than in e.g.
IC 3599 (M$_{BH}\approx 10^6$$M_{\odot}$).

\section{Discussion}

The nature of X-ray transient sources is of different origin. While in WPVS007
the question is why its X-ray source is off right now, the question for IC 3599
and RX J1624.9+7554 is what was the reason for the outburst. The X-ray flux seen in
WPVS007 during the RASS is in good agreement what is expected from the mean
optical to X-ray flux ratio (see \cite{beuermann1999}). A possible explanation
is a temperature change in the Comptonization layer above the disk that shifts
the UV photons into the soft X-ray range \cite{grupe1995b} and caused that
the observed spectrum is shifted out of the ROSAT PSPC energy window. 
The outburst in
IC3599 and RX J1624.9+7554 can be explained by accretion events, e.g. the tidal
disruption of a star that comes to close to the central black hole,
 a scenario that has been suggested
by e.g. \cite{rees1990}. Such X-ray outburst events are rare. However,  performing 
new soft X-ray surveys in order to find more of
these x-ray transient sources.  
Quick follow-up observations in the optical  and in X-rays would provide as
with a powerful tool to map the inner region of an AGN or galaxy while the light
front is passing through the inner region of the AGN.


\begin{thebibliography}{8.}
\addcontentsline{toc}{section}{References}

\bibitem {beuermann1999} Beuermann,K., Thomas, H.-C., Reinsch, K., et al., 1999, A\&A 347, 47 
\bibitem {brandt1995} Brandt, W.N., Pounds, K., Fink, H.H., 1995, MNRAS 273, L47 
\bibitem {greiner2000} Greiner, J., Schwarz, R., Zharikov, S., Orio, M., 2000, A\&A 362,
L25
\bibitem {grupe1995a} Grupe, D., Beuermann, K., Mannheim, K., et al., 1995a, A\&A 299, L5 
\bibitem {grupe1995b} Grupe, D., Beuermann, K., Mannheim, K., et al., 1995b, A\&A 300, L21 
\bibitem {grupe1999} Grupe, D., Thomas, H.-C., Leighly, K.M., 1999, A\&A 350, L31 
\bibitem {grupe2001a} Grupe, D., Thomas, H.-C., Beuermann, K., 2001a, A\&A 367, 470 
\bibitem {grupe2001b} Grupe, D., Thomas, H.-C., Leighly, K.M., 2001b, A\&A 369, 450 
\bibitem {komossa1999} Komossa \& Greiner, 1999, A\&A 349, L45 
\bibitem {kombade} Komossa, S., Bade, N., 1999, A\&A 343, 775
\bibitem {rees1990} Rees, 1990, Science 247, 817 
\bibitem {voges1999} Voges, W., Aschenbach, B., Boller, Th., et al., 1999, A\&A 349, 389 


\end{thebibliography}
\end{document}